\documentclass{aastex62}

\newcommand{\rep}[1]{{ #1}}
\newcommand{\repII}[1]{{ #1}}

\received{\today}
\revised{\today}
\accepted{\today}
\submitjournal{ApJ}

\shorttitle{Assessing an HII region on the far side of the Galaxy}
\shortauthors{Chen\'e et al.}

\begin{document}

\title{Assessing the stellar population and the environment of an HII region on the far side of the Galaxy\footnote{refers as VVV\,CL177}}

\correspondingauthor{Andr\'e-Nicolas Chen\'e}
\email{andrenicolas.chene@gmail.com}

\author[0000-0002-1115-6559]{Andr\'e-Nicolas Chen\'e}
\affil{Gemini Observatory/NSF’s NOIRLab, 670 N. A`ohoku Place, Hilo, Hawai`i, 96720, USA}
\collaboration{(VVV survey)}

\author{Robert A. Benjamin}
\affiliation{Department of Physics, University of Wisconsin-Whitewater, 800 West Main Street, Whitewater, WI 53190, USA}
\collaboration{(Glimpse survey)}

\author[0000-0002-7754-9785]{Sebastian Ram\'irez Alegr\'ia}
\affiliation{Centro de Astronom\'ia (CITEVA), Universidad de Antofagasta, Av. Angamos 601, Antofagasta, Chile.}
\author{Jura Borissova}
\affiliation{Instituto de F\'isica y Astronom\'ia, Universidad de Valpara\'iso, Av. Gran Breta\~na 1111, Playa Ancha, Casilla 5030, Valpara\'iso, Chile}
\affiliation{Instituto Milenio de Astrof\'isica, Santiago, Chile}
\author{Radostin Kurtev}
\affiliation{Instituto de F\'isica y Astronom\'ia, Universidad de Valpara\'iso, Av. Gran Breta\~na 1111, Playa Ancha, Casilla 5030, Valpara\'iso, Chile}
\affiliation{Instituto Milenio de Astrof\'isica, Santiago, Chile}
\author{Christian Moni Bidin}
\affiliation{Instituto de Astronom\'ia, Universidad Cat\'olica del Norte, Antofagasta, Chile}
\author{Francesco Mauro}
\affiliation{Instituto de Astronom\'ia, Universidad Cat\'olica del Norte, Antofagasta, Chile}
\author{Phil Lucas}
\affiliation{Centre for Astrophysics Research, University of Hertfordshire, College Lane, Hatfield, AL10 9AB, UK 0000-0002-8872-4462}
\author{Zhen Guo}
\affiliation{Dept. of Astronomy, University of Hertfordshire, Hertfordshire, UK 0000-0002-8872-4462}
\author{Leigh C. Smith}
\affiliation{Institute of Astronomy, University of Cambridge, Madingley Road, Cambridge CB3 0HA, UK}
\affiliation{Centre for Astrophysics Research, School of Physics, Astronomy and Mathematics, University of Hertfordshire, College Lane, Hatfield AL10 9AB, UK}
\author{Carlos Gonzalez-Fernandez}
\affiliation{Institute of Astronomy, University of Cambridge, Madingley Road, Cambridge CB3 0HA, UK}
\author{Valentin D. Ivanov}
\affiliation{European Southern Observatory, Karl-Schwarszchild-Str. 2, D-85748 Garching bei Muenchen, Germany 0000-0002-5963-1283}
\author{Dante Minniti}
\affiliation{Departamento de F\'isica, Facultad de Ciencias Exactas, Universidad Andr\'es Bello, Av. Fernandez Concha 700, Las Condes, Santiago, Chile}
\affiliation{Vatican Observatory, V00120 Vatican City State, Italy 0000-0002-7064-099X}
\collaboration{(VVV survey)}

\author[0000-0001-8800-1793]{L.D.~Anderson}
\affiliation{Department of Physics and Astronomy, West Virginia University, Morgantown, WV 26505, USA}
\affiliation{Center for Gravitational Waves and Cosmology, West Virginia University, Morgantown, Chestnut Ridge Research Building, Morgantown, WV 26505, USA}
\affiliation{Green Bank Observatory, P.O. Box 2, Green Bank, WV 24944, USA}
\author[0000-0002-7045-9277]{W.P.~Armentrout}
\affiliation{Green Bank Observatory, P.O. Box 2, Green Bank, WV 24944, USA}
\author{Danilo Gonzalez}
\affiliation{Instituto de Astronom\'ia, Universidad Cat\'olica del Norte, Antofagasta, Chile}
\affiliation{Instituto de F\'isica, Universidad de Antioquia, Calle 70 52-21, Medellin, Colombia}
\author{Artemio Herrero}
\affiliation{Instituto de Astrof\'isica de Canarias, 38200, La Laguna, Tenerife, Spain}
\affiliation{Departamento de Astrof\'isica, Universidad de La Laguna, 38205, La Laguna, Tenerife, Spain}

\author{Karla Pe\~na Ram\'irez}
\affiliation{Centro de Astronom\'ia, Universidad de Antofagasta, Antofagasta, Chile}
\nocollaboration



\begin{abstract}

We have investigated the stellar and interstellar content of the distant star formation region \object{IRAS 17591-2228} (WISE HII region \object{GAL 007.47+0.06}). It is associated to a water maser, whose parallax distance is $d=20.4^{+2.8}_{-2.2}$\,kpc, supported by independent measurements of proper motion and radial velocity. It is projected in the same direction as an extremely red ($J$$-$$K_S$\,$\sim$\,6\,mag) group of stars, and a shell of mid-infrared emission.
\repII{We qualify the group of stars as a cluster candidate, VVV\,CL177. Its radius spans between 0.45$\arcmin$ and 1$\arcmin$ and contains at least two young stellar objects with an extreme extinction near A$_{\rm V}\sim40$\,mag. Yet more analysis will be required to determine is it is a real single cluster associated with the water maser. The $^{13}$CO emissions at the radial velocity of the maser corresponds to the mid-infrared emission. }

\end{abstract}



\section{Introduction}\label{sec:intro}

The HII region \object{GAL 007.47+0.06} was discovered using the Effelsberg 100-m radio telescope by \citet{wink82} who reported two possible heliocentric distances of 6.3 and 25.1 kpc. Using direct trigonometric parallax of a water maser source associated with \object{GAL 007.47+00.06} \citep{Lo89}, \citet{Sa17} estimated $d=20.4^{+2.8}_{-2.2}$\,kpc, placing it in the Outer Scutum-Centaurus spiral arm as it passes through the far side of the Milky Way (MW). They determine its radial velocity (RV) with V$_{\rm LSR}=$-16$\pm$4 km\,s$^{-1}$, which is consistent with values obtained from a number of different tracers \citep[e.g.][]{Br96,balser11}. Though, \citet{yamauchi16} suggests the possibility of another cloud being in front of \object{GAL 007.47+00.06} with V$_{\rm LSR}=$+15 km\,s$^{-1}$.

The direction where this active formation region is located is hard to study as the line of sight crosses most of the Milky Way and suffers extreme extinction. Nevertheless, the maser source is within a few arcseconds of the center of a bright 1.50$\arcmin$-wide mid-infrared nebula (seen in the GLIMPSE II survey, \citealt{Benjamin03,Churchwell09}, at 8.0\,$\mu$m, and a fairly compact group of about a dozen stars that share the same highly reddened color, seen in the VISTA Variables in the V\'ia L\'actea survey \citep[VVV,][]{Mi10,Sa12,He14}. The mid-infrared nebula is also seen in the WISE $W3$ and $W4$ passbands \citet{Wr10}.

In this paper \repII{we study the stars, gas and dust near the central coordinates of \object{GAL 007.47+00.06}, and verify if they are likely related to one another and to the HII region and its water maser.}

\begin{figure}[htbp]
    \centering
    \includegraphics[angle=0,width=2.3in]{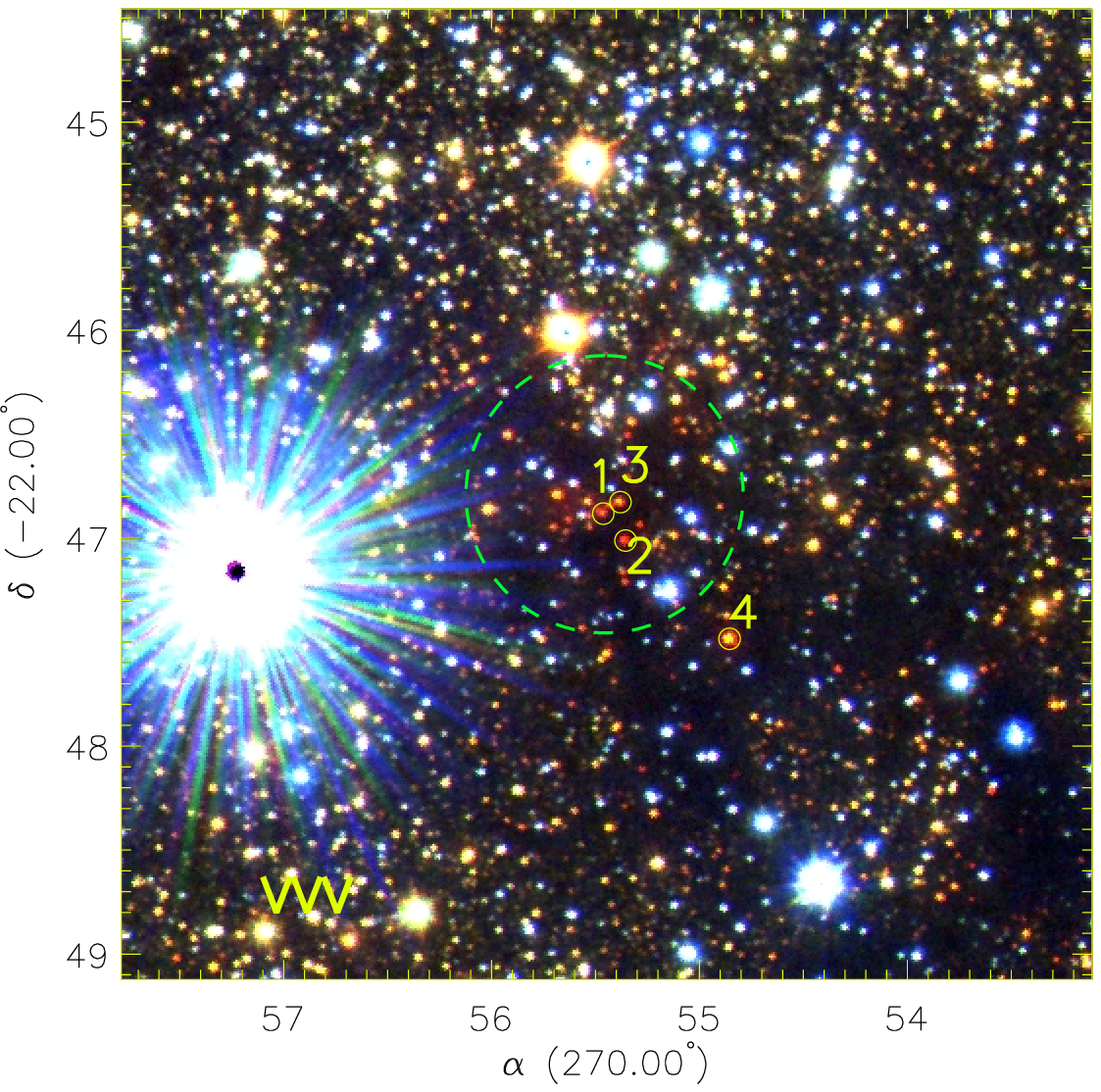}
    \includegraphics[angle=0,width=2.3in]{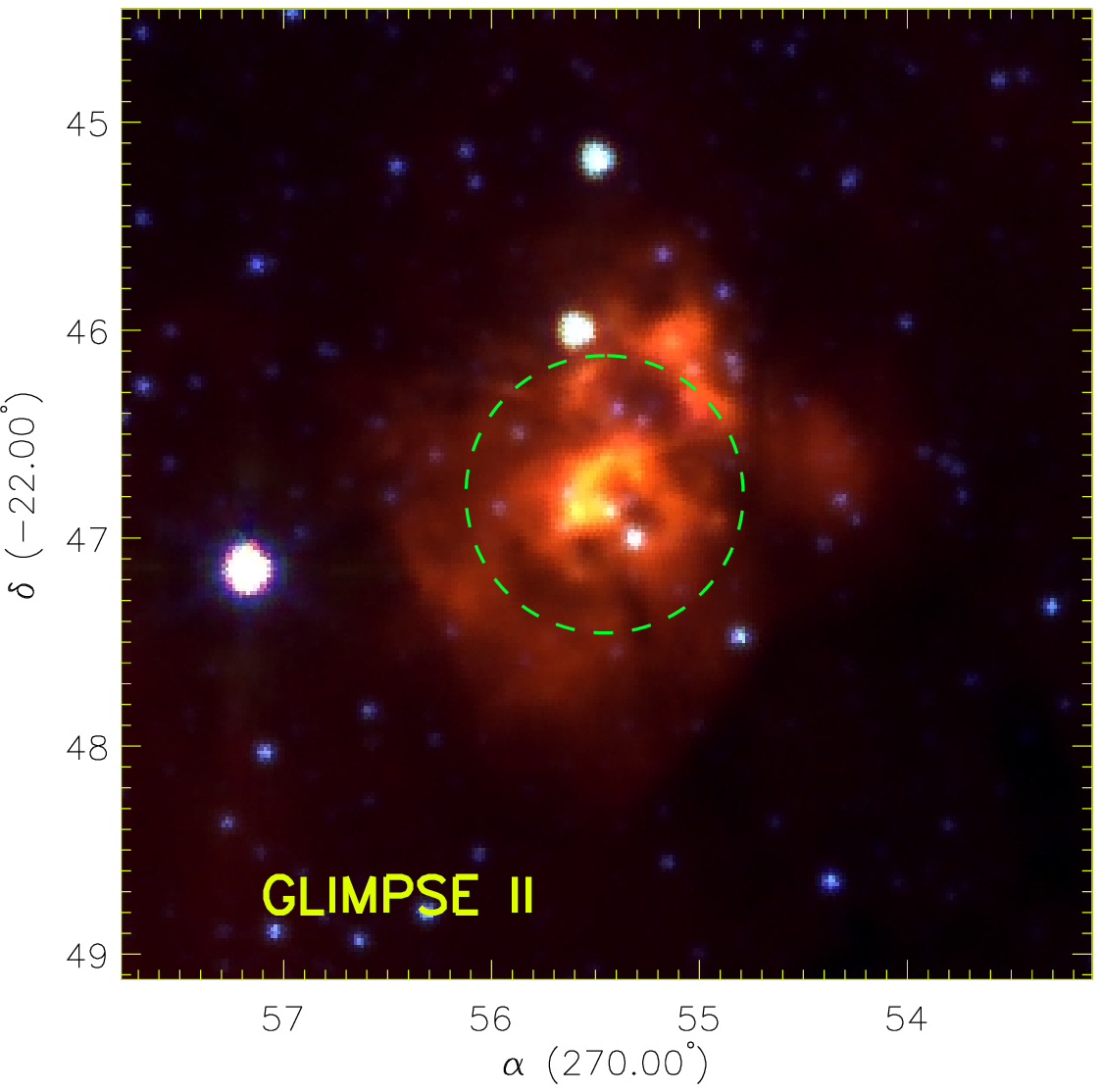}
    \includegraphics[angle=0,width=2.3in]{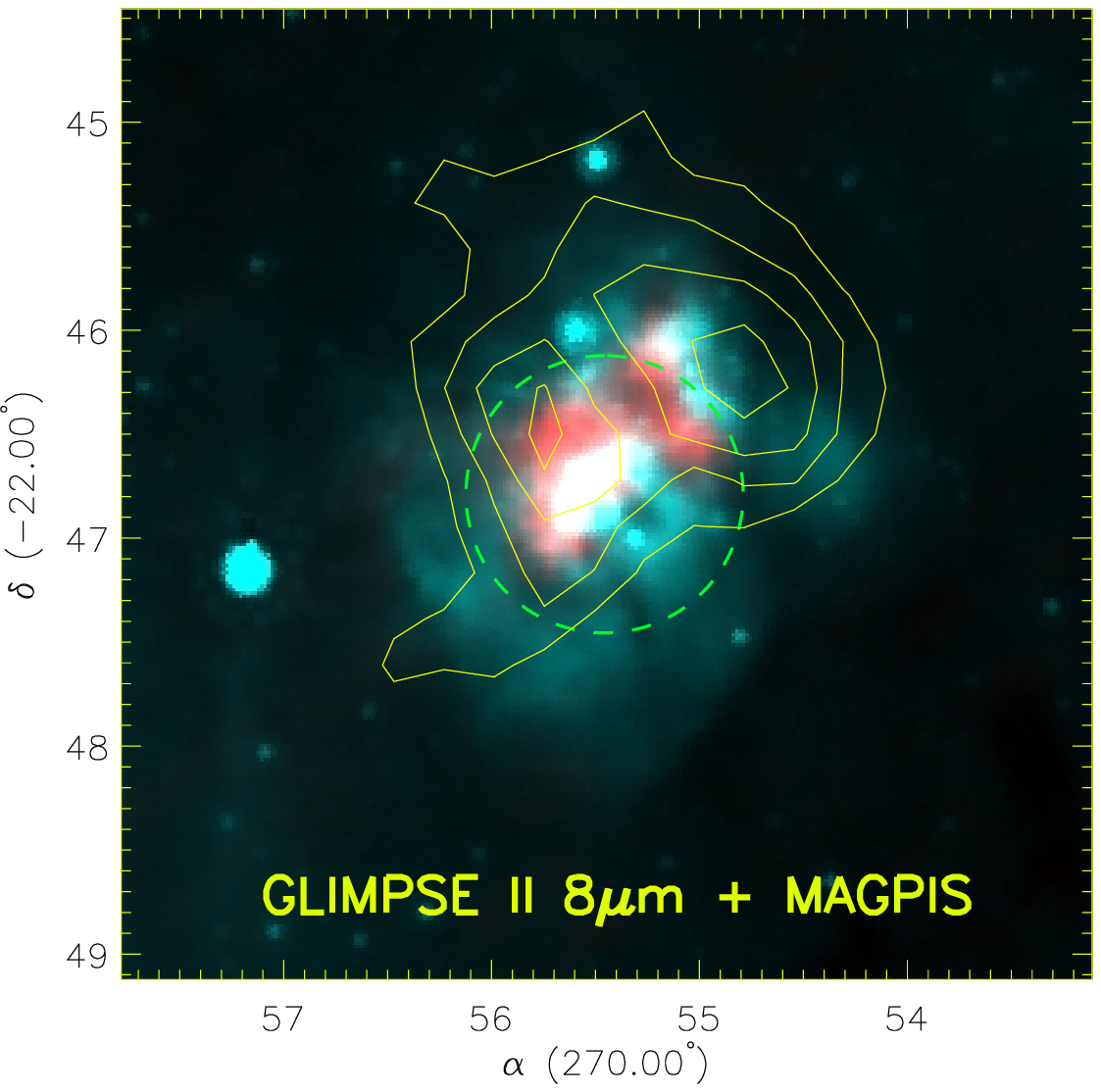}
    \caption{(Left): $JHK_S$ false-colour images of the \object{GAL 007.47+0.06} region. Stars labelled with yellow numbers were observed using NIR spectrographs. The green dashed circle indicates the angular sizes of the cluster candidate. (Middle): Three-colour 4.5\,$\mu$m (blue), 5.8\,$\mu$m (green) and 8\,$\mu$m (red) GLIMPSE II images of the region of \object{GAL 007.47+0.06}. (Right): Combination of the GLIMPSE II 8\,$\mu$m emission (turquoise) and the MAGPIS 20-cm radio continuum emission from the ionized gas (red). Directions in which both the 8\,$\mu$m and radio emission are present appear in white, independent of the intensity of the emission. Contours of $^{13}$CO emission from the Green Bank Telescope are plotted with yellow solid lines. This $^{13}$CO emission \repII{was obtained by integrating only between $-$20 km s$^{-1}$ and $-$10 km s$^{-1}$ (see Section~\ref{sec:radio} for more details)}. All three panels are centered on the water maser coordinates \citep[][]{Sa17}, and have a 4.5$\arcmin\times$4.5$\arcmin$ area. North is up, East is to the left, and coordinates are given in the J2000 system.}
      \label{FOV}
\end{figure}

\newpage
\section{observations}\label{Obs}
We used near infrared (NIR) images and photometric data from the  ESO Public Survey VVV, which observes with the VISTA InfraRed CAMera (VIRCAM) at the VISTA 4 m telescope at Paranal Observatory \citep{Em10} and reduced at CASU\footnote{http://casu.ast.cam.ac.uk/} using the VIRCAM pipeline v1.3 (Irwin et al. 2004). A $JHK_S$ false-colour image of the cluster is shown in Fig.\,\ref{FOV}. Stellar photometry was performed by employing the VVV-SkZ\_pipeline's \citep{Ma13} automated software based on ALLFRAME \citep{St94}, optimized for VISTA point-spread function photometry. 2MASS photometry was used for absolute flux calibration in the $J$, $H$, and $K_S$-bands, using stars with 12.5$<$$J$$<$14.5\,mag, 11.5$<$$H$$<$13\,mag, and 11$<$$K_S$$<$12.5\,mag. 

The mid infrared (MIR) regime was studied using images and photometric data from the GLIMPSE II survey. The images and data were downloaded from the NASA/IPAC Infrared Science Archive\footnote{https://irsa.ipac.caltech.edu/data/SPITZER/GLIMPSE/doc/glimpse2\_dataprod\_v2.1.pdf}.

We use 1.4\,GHz Multi-Array Galactic Plane Imaging Survey (MAGPIS) data from \citet[][]{He06}. MAGPIS has an angular resolution of $\sim\!5\arcsec$, and has good sensitivity to extended emission.

\begin{figure}[htbp]
    \centering
    \plotone{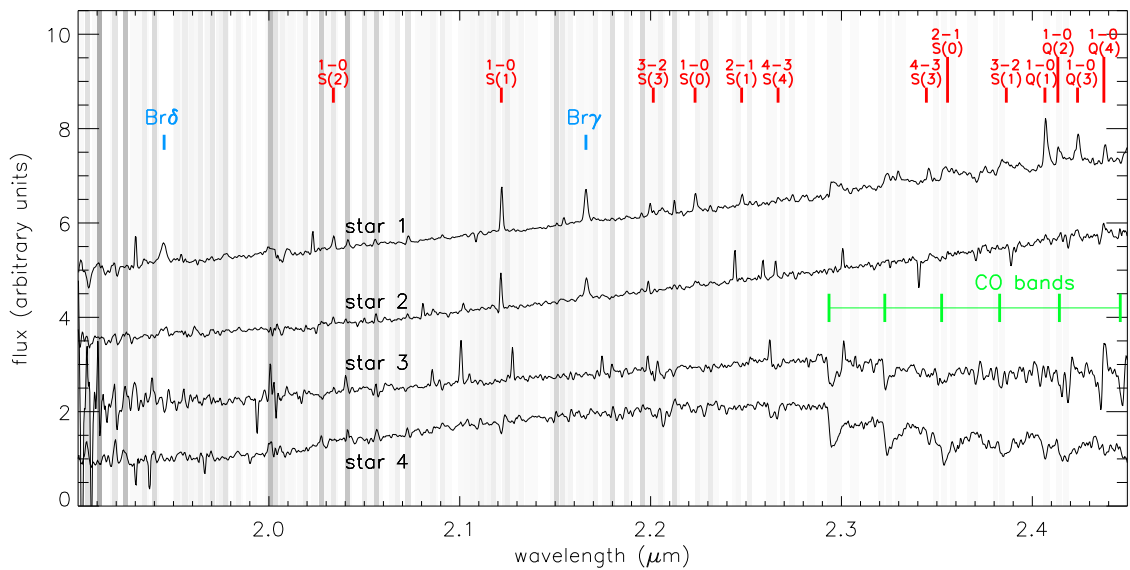}
    \caption{Spectra of four stars in the vicinity of the water maser. The H recombination lines, H$_2$ lines and CO-band features are marked with blue, red and green labels, respectively. Gray vertical bands indicate the location of OH emission lines. The darker the band, the stronger the line.}
    \label{sp}
\end{figure}

We collected spectra of the brightest red stars within the cluster's radius using Flamingos-2 (F2) at the Gemini Observatory\footnote{Program ID GS-2018A-FT-209} (see Fig.\,\ref{sp}). These stars are labeled with yellow numbers in Fig.\,\ref{FOV} (left). The resolution power is between 3000 and 4000, and the wavelength range covers only the $K$ band. For optimal subtraction of the atmospheric OH emission lines, we used nodding along the slit. The average signal-to-noise ratio (S/N) per pixel ranges from 50 to 150. Bright stars of spectral type B8 to A2 were observed as a measure of the atmospheric absorption and selected to share the same airmass as the targeted cluster stars during the middle of their observation. All reduction steps were executed with standard {\sc iraf}\footnote{{\sc iraf} was distributed by the National Optical Astronomy Observatories (NOAO), which is operated by the Association of Universities for Research in Astronomy, Inc. (AURA) under cooperative agreement with the U.S. National Science Foundation (NSF).} procedures via the Gemini {\sc iraf} package.

\section{Physical properties of the cluster candidate}\label{Descriptions}
The first assumption of this study is that the stars that share the same highly reddened color and are grouped a few arcseconds from the position of the water maser presented by \citet{Sa17} could potentially be members of a single cluster that is related with the maser at a distance of $\sim 20$\,kpc, as given by direct trigonometric parallax. \repII{This group of sources are surrounded by strong nebular 8\,$\mu$m-emission, reinforcing our hypothesis. It is examined in further in the next sub-sections. The goal of this section is to test this assumption with our data. The working name of the cluster candidate is VVV\,CL177, following the VVV cluster naming system.}

\begin{figure}[htbp]
    \centering
    \plottwo{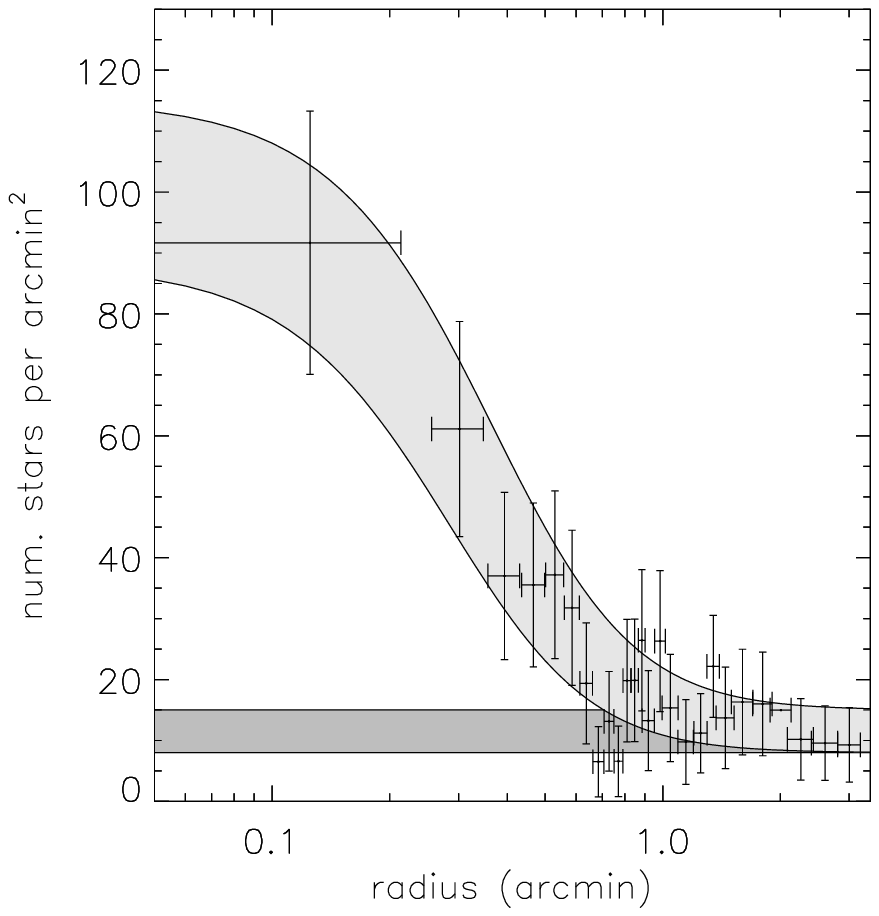}{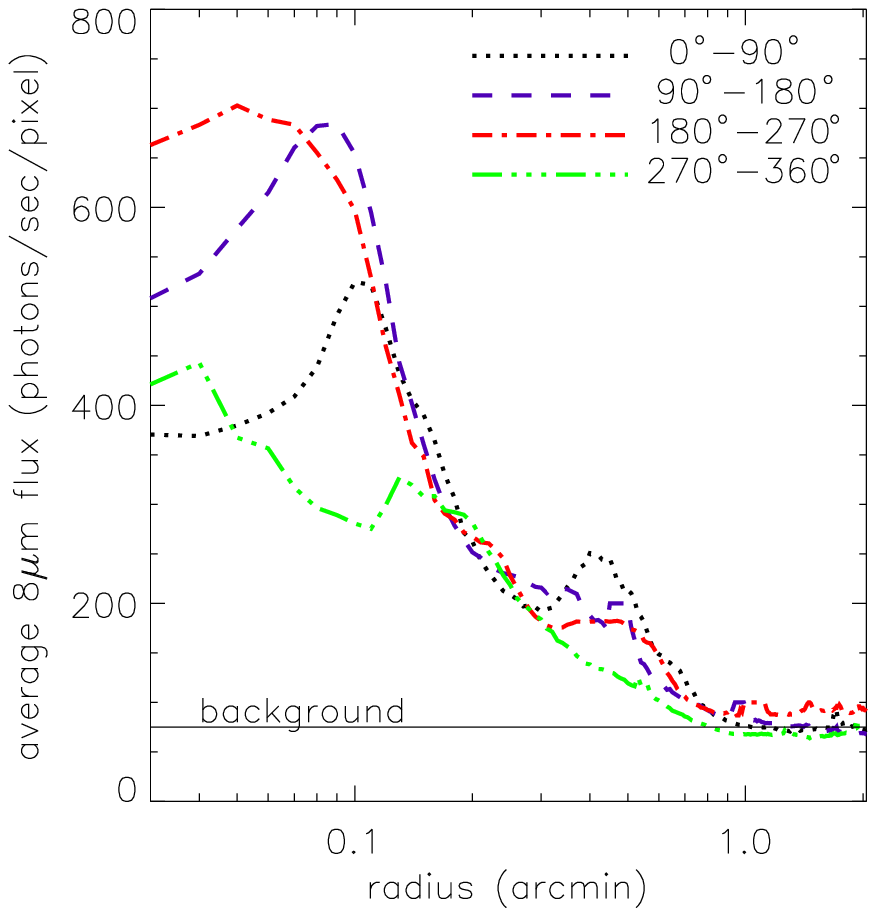}
    \caption{(Left): Radial density profile of the stars with colors ($H$$-$$K_S$)$>$2.4\,mag, centered on the coordinates of the water maser. The darker horizontal band represents the background level (the width is determined by the uncertainty on the value) and King profiles are given with uncertainties in light grey. (Right): Average flux measured in the 4 quadrants of the 8\,$\mu$m-emission GLIMPSE II image centered on the maser coordinates as a function of radius. Quadrant I (0$^\circ$ - 90$^\circ$) is plotted with a black dotted line, quadrant II (90$^\circ$ - 180$^\circ$) with a blue dashed line, quadrant III (180$^\circ$ - 270$^\circ$) with a red dashed-dotted line, quadrant IV (270$^\circ$ - 360$^\circ$) with a green dashed-triple dotted line.}
    \label{Rad}
\end{figure}
\subsection{The radius} \label{radius}
Assuming \repII{that VVV\,CL177 is a real cluster}, we determine its angular size from the radial density profile (RDP) based on our VVV stellar catalogue, using the coordinates of the water maser as the central coordinates. To increase the contrast between the density of the cluster \repII{candidate} and the background level, we exclude the stars with VVV ($H$$- $$K_S<2.4$)~mag from the catalogue, as they are more likely to be field stars, judging by the color and magnitude of stars more than 2$\arcmin$ away from the center of the cluster candidate. The result is presented in Fig.\,\ref{Rad}. The darker horizontal band marks the background level with its uncertainty. To better guide the determination of the \repII{VVV\,CL177}'s angular size, we fit a two-parameter King profile adapted to star counts \citep[as in][but using stars count instead of surface brightness]{Ki66}. The fit obtained and its uncertainty are plotted in lighter grey. The radius derived by the King profile fitting is 0.45$\arcmin$. Of course, such a cluster \repII{would} not be isolated, self-gravitating and in equilibrium, as required for use of the King profile, yet this value can be used as a first estimate. As comparison, the RDP seems to reach the background level between 0.7$\arcmin$ and 1.05$\arcmin$.

Interestingly, the RDP can be compared with the profile of the nebular emission. In Fig.\,\ref{Rad}, we present the average flux measured in the 4 quadrants of the 8\,$\mu$m-emission GLIMPSE II image centered on the maser coordinates as a function of radius. The morphology can vary significantly between quadrants, as they show a strong central cavity and local clumps, most  likely sculpted by strong stellar winds. Yet, all the profiles reach the background level at the same radius, which is close to 0.8$\arcmin$, and the nebula remains mostly spherical, centered on the water maser.

Both the RDP and the radial 8\,$\mu$m-emission profile seem to correspond, as if the density of stars (with the selected colors) is corresponding to the density of nebular gas. Without additional information, this observation could be explained in two possible ways. Either the nebula is in the foreground, reddening the stars behind it and giving the impression of a cluster of stars with comparable colors, or both the nebula and the stars with a  ($H$$-$$K_S$) colour greater than 2.4\,mag are related to the water maser, and are part of a giant star forming region. 

In the following sections, we will \repII{verify which of these scenarios is the most likely.} Assuming \repII{it is the latter, and using} the distance of $\sim$20\,kpc from the water maser parallax, the cluster diameter would be at least a minimum of 5\,pc, but potentially up to 9-10\,pc if one includes all the 8\,$\mu$m emission, and considers the farthest radius where the RDP reaches the background. \repII{In that case, }the nebula is only visible from 8\,$\mu$m and redder due to extreme reddening along the line of sight.

\begin{figure}[htbp]
    \centering
    \includegraphics[angle=0,width=3.25in]{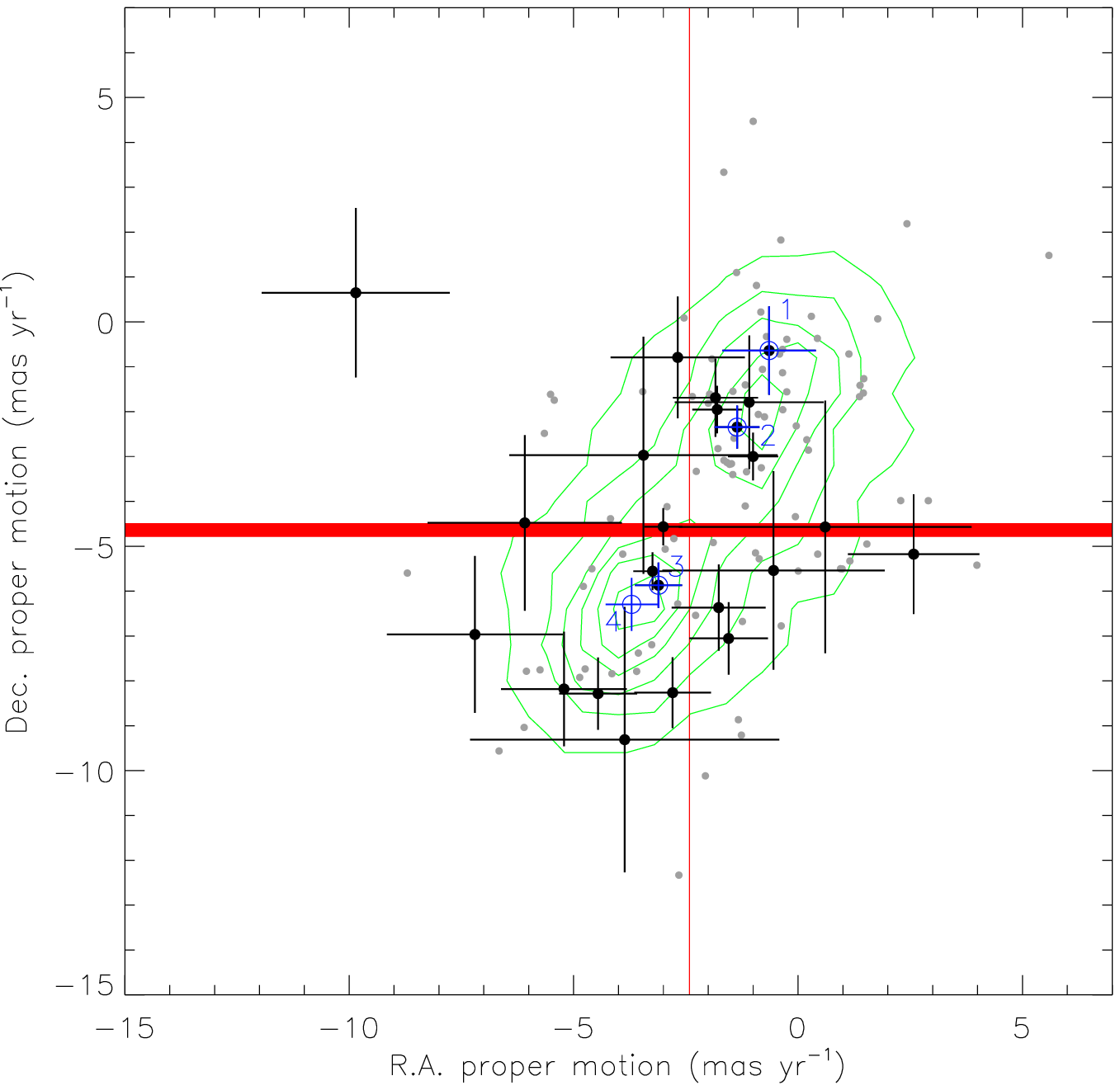}
    \includegraphics[angle=0,width=3.6in]{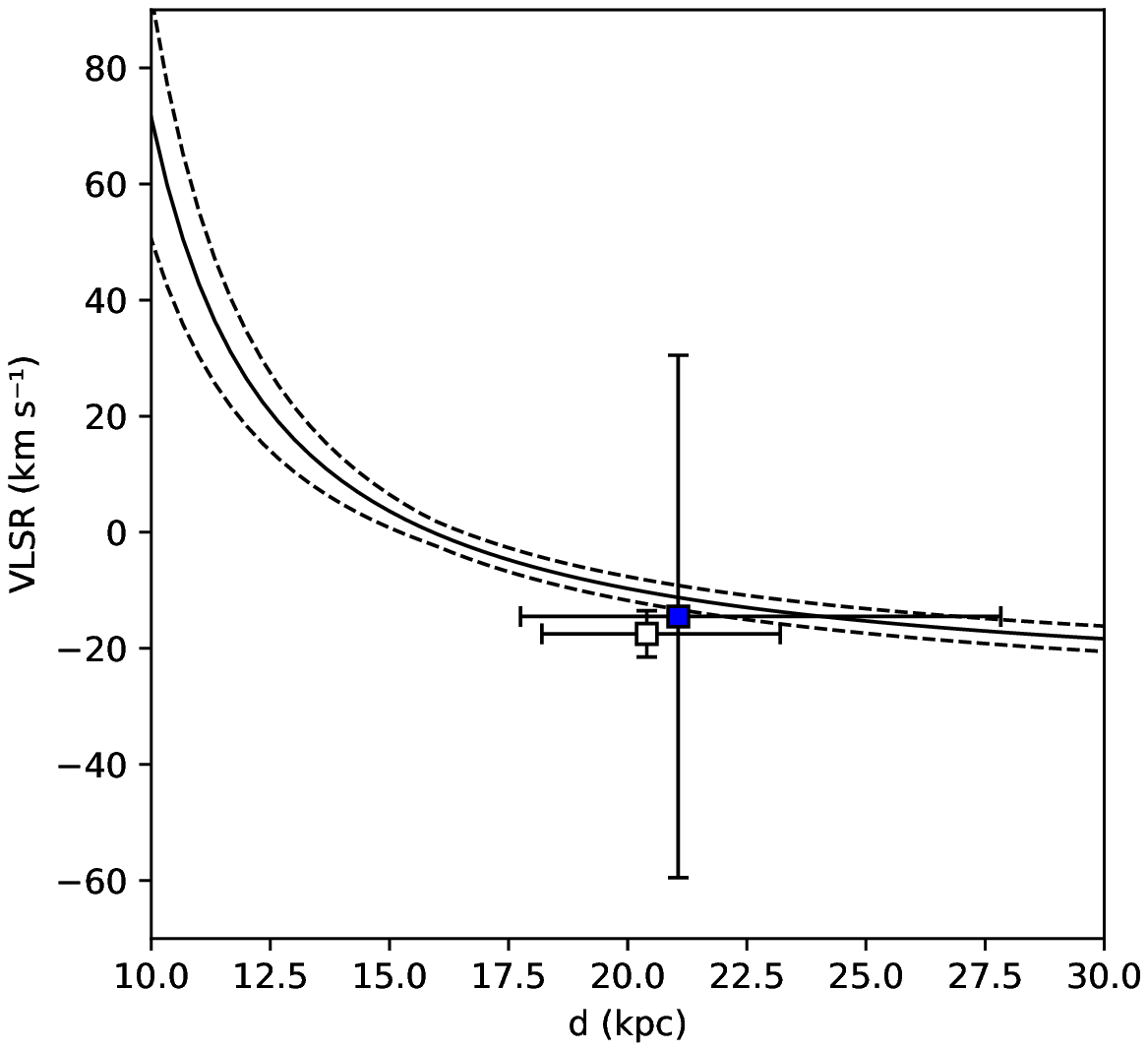}
    \caption{(Left): Proper motions (PMs) of stars within a 0.45$\arcmin$ radius around the water maser. The PM of the water maser is marked with red bars \rep{as wide as} a 1$\sigma$ error. \rep{The green contours show the distribution of the PMs of the stars within the radii 1.5$\arcmin$ and 4$\arcmin$ around the water maser coordinates. The lowest level is 45 stars/(mas$^2$yr$^{-2}$) and the highest is 275 stars/(mas$^2$yr$^{-2}$).} Stars with a color ($H$$-$$K_S$)$<$2.4\,mag are plotted in gray. Stars with a color ($H$$-$$K_S$)$>$2.4\,mag are either plotted in black. The uncertainties for the stars 1, 2 and 3 are plotted in blue. Star\,4 falls outside the 0.45$\arcmin$ radius, but was added using an empty blue circle. (Right):  \rep{Distance-RV profile of the Galaxy in the direction of \object{GAL 007.47+0.06} (solid curve). The profile was obtained assuming the rotation curve of \citet{Br93}, and dashed curves used to indicate the 1$\sigma$ propagation errors. The $\rm{V_{LSR}}$ for the water maser is $-$16$\pm$4\,km\,s$^{-1}$ (white square), and $-13\pm45$\,km\,s$^{-1}$ for star\,1 (blue square).} }
    \label{motion}
\end{figure}

\subsection{Membership}
In order to verify if the stars that share the same highly reddened color within the cluster candidate radius are bound and associated to the water maser, or not, we inspect their proper motions (PMs) distribution. We use the VVV InfraRed Astrometric Catalogue version 2 data \citep[VIRAC2,][section 4.1]{Sm18}.\repII{ We apply the Gaussian Mixture Model \citep[GMM;][ e.g. ]{Ev11} technique, which is based on the assumption that the stars distribution within an overdensity can be described by a superposition of multivariate gaussian distributions \citep[][]{dS17}. The GMM was implemented in R\footnote{https://www.r-project.org} using the mclust library \citep[][]{Sc16}. The mixture model includes a noise component for the field stars, which is initialized using the nearest neighbor cleaning method from \citet[][]{BR98} via the NNclean function in the prabcus library of R. We apply the GMM technique in the proper motion space, using only the stars within the cluster candidate radius. We use ten multivariate Gaussian components. The results are insensitive to the exact number of Gaussians, as long as it is large enough to disentangle the background from the cluster candidate members. The two most populated Gaussians (44 and 24 objects each) have a considerable mixing probabilities ($<$ 28\%). Nevertheless, their proper motion values ($\mu_{\rm RA_1}=-1.28\pm0.29$\,mas\,yr$^{-1}$, $\mu_{\rm DEC_1}=-2.05\pm0.43$\,mas\,yr$^{-1}$, $\mu_{\rm RA_2}=-3.01\pm0.44$\,mas\,yr$^{-1}$, $\mu_{\rm DEC_2}=-6.87\pm0.33$\,mas\,yr$^{-1}$ are between 3 and 6\,$\sigma$ from the water maser PMs \citep[$\mu_{\rm RA}=-2.43\pm0.1$\,mas\,yr$^{-1}$, $\mu_{\rm DEC}=-4.43\pm0.16$\,mas\,yr$^{-1}$,][]{Sa17}.

This result remains the same when we select the stars within an annulus around the water maser coordinates. As can be seen in Fig.\,\ref{motion} (left), the} distribution of PMs, plotted in green contours, shows two groups: one with PMs more negative than the maser, and the other with PMs more positive. The stars within the cluster candidate radius (0.45\arcmin) are plotted with gray or black dots, depending on their color ($H$$-$$K_S$) being lower or higher than 2.4\,mag, respectively. The distribution of the redder stars PMs is surrounding the water maser PM, but also mimics the reference field. Though, note that there are potential systematic issues in the transformation of the PMs into the Gaia system, and errors may be underestimated. 

Our $K-$band spectra in Fig.\,\ref{sp} cover 4 stars with a ($H$$-$$K_S$) color greater than 2.4\,mag. These spectra present three major features: H recombination emission lines (Br$\delta$ and Br$\gamma$ at 1.95\,$\mu m$ and 2.16\,$\mu m$, respectively), H$_2$ lines in emission and the ${}^{12}$CO $\Delta \nu = 2$ bandheads (${}^{12}$CO (2,0) at 2.29, ${}^{12}$CO (3,1) at 2.32,  and ${}^{12}$CO (4,2) at 2.35\,$\mu m$).
 
\rep{Star\,2 shows only the H recombination and H$_2$ lines. The stars 3 and 4 show CO-bands in absorption and with a depth similar to the observed in giant stars.} By contrast, star\,1 shows the CO-bandheads clearly in emission, features related to accretion by young stellar objects \citep[YSOs][]{Sc79}. This spectrum confirms that at least one of the stars is an accreting YSO and \repII{could} be associated with the star forming region at the source of the maser emission. The spectra of stars 1 and 2 are typical of fairly massive embedded YSOs in the literature \citep{cooper13}. 

\repII{Note that for star\,1, the line ratios of 1-0 S(2), 2-1 S(3), 2-1 S(2), 3-2 S(3), and 1-0 S(0) to 1-0 S(1) give $0.45\pm0.06$, $0.19\pm0.01$, $0.17\pm0.01$, $0.09\pm0.01$ and $0.49\pm0.07$, respectively, once corrected for extinction (determined in the next section), which has an effect on the same scale as the uncertainties. Such ratios correspond to an UV fluorescence excitation mechanism \citep{Wo91}. }

\rep{The spectral resolution and the signal-to-noise ratio of the spectra do not allow for radial velocity (RV) measurements more precise than 40\,km\,s$^{-1}$. Using the ${}^{12}$CO bandheads observed in stars\,1, 3 and 4, we find $\rm{V_{LSR}}=-13\pm$45\,km\,s$^{-1}$, $-249\pm$40\,km\,s$^{-1}$ and $-155\pm$40\,km\,s$^{-1}$, respectively. While the RV for star\,1 is in agreement with $\rm{V_{LSR}}=-16\pm$4\,km\,s$^{-1}$ for the water maser associated with \object{GAL 007.47+0.06} \citep{Sa17}, stars 3 and 4 are off by a significant factor, ruling out their membership of the cluster completely. Note that no RV could be obtained for star\,2, as it only displays narrow emission lines. Those cannot be trusted, because the H recombination lines (Br$\delta$ and Br$\gamma$) \rep{typically arise either in a fast wind or gas falling on to the star at free-fall velocity}, and the H$_2$ lines typically arises in an outflow or wind that is often offset from the systemic velocity \citep{guo2020}. 

The Fig.\,\ref{motion} (right) shows star\,1 $\rm{V_{LSR}}$ (blue square) compared to that of the maser (white square). The distance estimate to star\,1 is $21.06_{-3.31}^{+6.77}$ kpc, and determined following the method described in \citet[][section 4]{Re09}, using the Galactic parameters from the model A5 in \citet{Re14}.}

\begin{figure}[htbp]
    \centering
    \includegraphics[angle=0,width=3in]{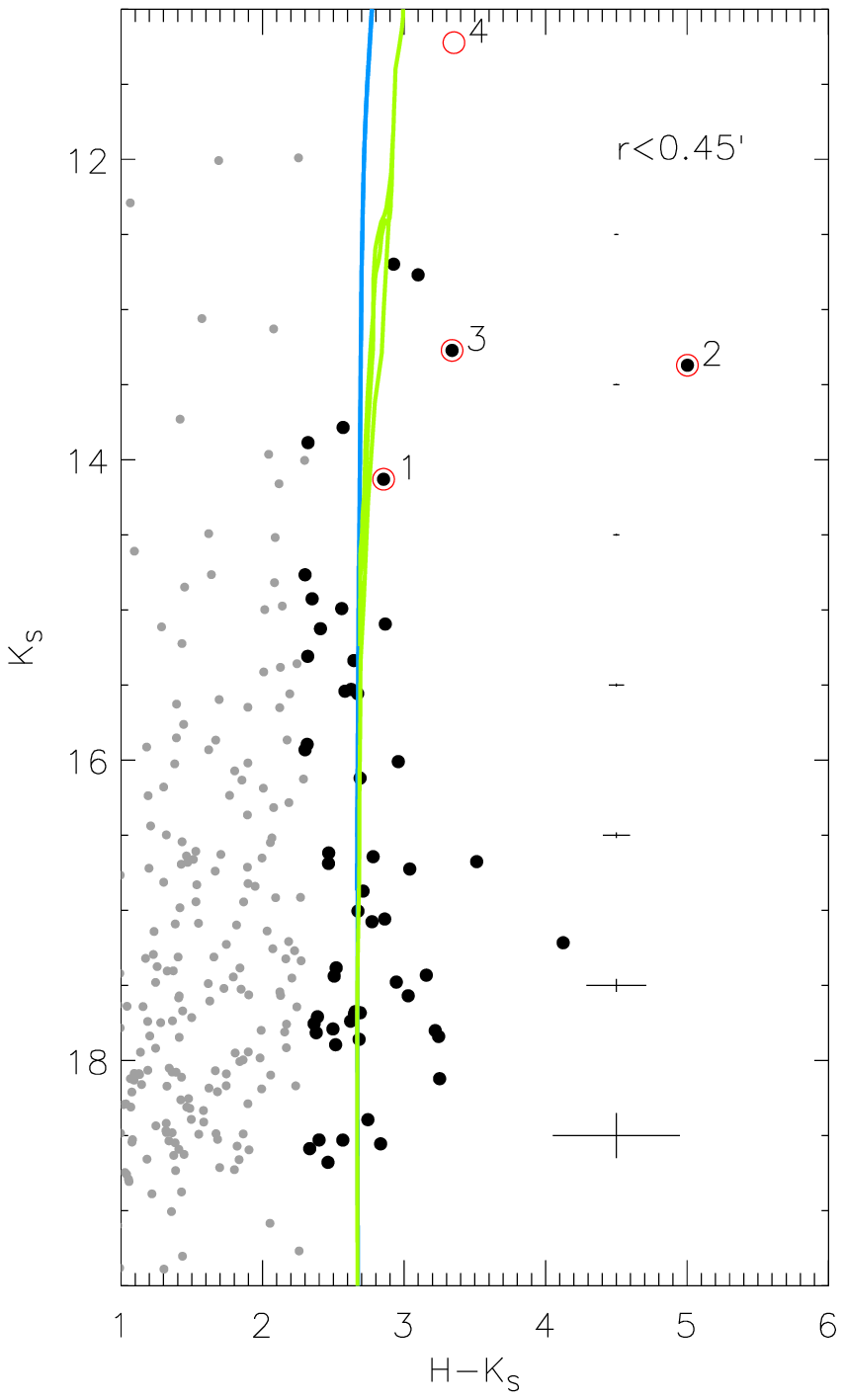}
    \includegraphics[angle=0,width=3in]{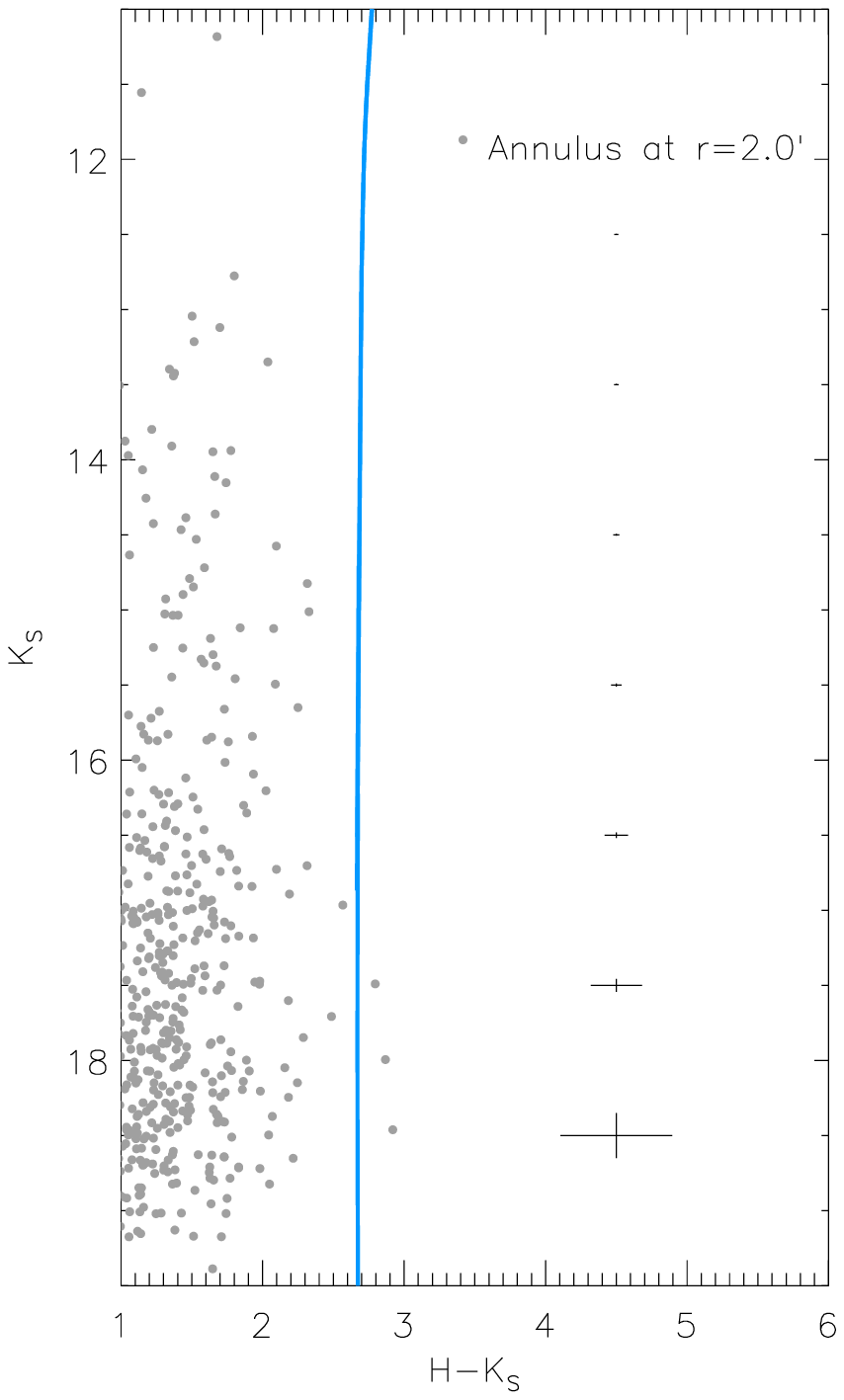}
    \caption{(Left): ($H$$-$$K_S$) vs. $K_S$ color-magnitude diagram of the stars within a radius of $r=0.45\arcmin$ from the center of the cluster candidate. The average photometric errors are plotted for each $K_S$ magnitude. Isochrone from \citet{Ek12} are plotted for 1 (blue) and 20 Myr (green). \rep{Stars 1 to 3 are circled in red, and star\,4 is added (as an empty circle), even though it is outside the cluster radius.} (Right): Same, but for an annulus around the cluster candidate, centered at $r=2\arcmin$, and covering the same area as for the diagram of the cluster candidate. The 1 Myr isochrone is plotted for reference.}
    \label{fig:cmd}
\end{figure}

\repII{In summary, while at least some stars are YSOs with an RV that does not exclude a distance of 20\,kpc, there are no indications based on PMs that the stars that share the same highly reddened color and are grouped a few arcseconds from the position of the water maser are members of a unique cluster.}

\subsection{Age and extinction} \label{sec:age_ext}

\repII{Even if the PMs do not support the hypothesis that all the reddest stars within VVV\,CL177 estimated radius are members of a unique cluster, we can still assume that some of the stars may be member of a cluster associated to the water maser.

To determine the value for the extinction, we exploit the ratio of 1–0 S(1) and Q(3) emission lines of H$_2$. Since Q(3)/S(1) = 0.70 from the spontaneous decay rates \citep{Tu77,Wo98}, any deviation from that value is due to reddening, as it is affecting the S(1) more than the Q(3) line. For star\,1, the star most likely associated with the water maser, Q(3)/S(1)\,=\,1.7$\pm$0.1, which, using the equations from \citet[][]{OC05} corresponds to A$_{\rm V}$=38.0$\pm$2.5\,mag. }

The Fig.\,\ref{fig:cmd} (left) shows the colour-magnitude diagram (CMD) $H$$-$$K_S$ as a function of $K_S$ of the cluster within the estimated radius. Stars 1 to 4 from Fig.\,\ref{sp} are marked. \repII{With no trustworthy membership information available, we simply apply the same strict cut at a color of $H$$-$$K_S$=2.4\,mag as in Section\,\ref{radius}, giving a rough first look at what VVV\,CL177's CMD could look like}. There is no detected flux in the $J$-band \rep{for most of the stars, as} the brightest star barely reaches $J$$\sim$18\,mag (corresponding to a photometric error greater than 0.1\,mag in VVV). 


Using the water maser parallax distance \repII{and our value for extinction, we plot isochrones of young clusters from \citet{Ek12} over the CMD on Fig.\,\ref{fig:cmd} (left). Interestingly, the position of the isochrones corresponds to the color cut used to select the stars representing the putative cluster, even if we now know that most of them may not be members of VVV\,CL177 according to their PMs. Also, there are no stars near those isochrones in the CMD obtained using an annulus at a 2$\arcmin$ radius from the center of VVV\,CL177 as comparison field (Fig.\,\ref{fig:cmd}, right). Based on this first estimate, would VVV\,CL177 be a real cluster, it would be younger than 20\,Myr old, and only the most massive stars would visible in our 
data.} 

\subsection{Radio luminosity and total molecular gas mass}
\label{sec:radio}
\repII{We mapped the $^{13}$CO emission with the Argus instrument on the 100 meter Green Bank Telescope. Argus is a 16-pixel focal plane array operating between frequencies of 74--116 GHz, designed for fast mapping of molecular gas. While there were several $^{13}$CO components along the line of sight, there was an isolated line between $-$20 km s$^{-1}$ and $-$10 km s$^{-1}$ associated with the LSR maser velocity of $-$16 $\pm$ 4 km s$^{-1}$ from \citet{Sa17}. We integrated only along this isolated $^{13}$CO component to obtain the map displayed in Fig.\,\ref{FOV} (right), and the flux distribution is corresponding to the position of the 8$\mu m$ nebula. There is some offset, of course, since the $^{13}$CO emission traces cold gas that has not yet formed into stars. Such a match indicated that both the cold molecular gas and the nebula are situated at the same distance as the water maser. For the rest of this section, we will adopt a distance of $\sim$20\,kpc.}

 We assume a $^{13}$CO-to-H$_{2}$ molecular gas conversion factor of X$_{^{13}\mathrm{CO}}$ = 10$\times$10$^{20}$ cm$^{-2}$ (K\,km\,s$^{-1}$)$^{-1}$ \citep{cormier18} which corresponds to a $^{13}$CO mass conversion factor, $\alpha_{^{13}\mathrm{CO}}$, of 21.62. Using this conversion factor, our integrated intensity (W$_{^{13}\mathrm{CO}}$) of 3.26 K km s$^{-1}$, an assumed distance (D) of 20 kpc, and the solid angle of the source ($\Omega_\mathrm{S}$) of 5.44$\times$10$^{-8}$ Sr, we can estimate a total molecular gas mass from \citet{bolatto13}: M$_{\mathrm{mol}}[\mathrm{H}_2] = \alpha_{^{13}\mathrm{CO}}~{\rm L}_{^{13}\mathrm{CO}} = \alpha_{^{13}\mathrm{CO}}~W_{^{13}\mathrm{CO}}~D^2~\Omega_{\mathrm{S}}$.
This results in a total molecular gas mass of 1.60$\times$10$^3$ M$_{\odot}$. Separately, the {\it Herschel}/HiGal survey detected the far infrared emission of the dust in the associated molecular clump \citep{elia17}. Unsure of the distance at that time, the survey team calculated a nominal mass of 36.72~M$_{\odot}$ at a standardised distance of 1~kpc. Scaling their nominal mass to the maser parallax distance, we find M$_{\mathrm{mol}}[\mathrm{H}_2]$ = 1.50$\times$10$^4$ M$_{\odot}$.

\repII{Using MAGPIS 1.4\,GHz radio continuum data we find a flux density of 1.8\,Jy, which at the assumed distance} is equivalent to a Lyman continuum photon production rate of 6.5$\times$10$^{49}$\,s$^{-1}$ \citep{Rub68}, assuming an electron temperature of 8000\,K. This Lyman continuum luminosity is more than the output of a single main sequence O3 star \citep{Mar05} and makes the region one of the most luminous in the Milky Way. The spatial scale is approximately 3~pc, i.e. a classical HII region rather than a UC HII region. It is more luminous than any of the $\sim\!1000$ first-quadrant HII regions regions in \citet{Ma17} or any of the 213 UC HII regions in \citet{Ur13} though the G055.114+02.422 region \citep{armentrout17} is almost as luminous and located at an even larger Galactocentric radius, also in the Outer Scutum-Centaurus arm. Furthermore, the radio flux density is likely a lower limit because given the molecular gas morphology, many photons may be leaking out of the region and therefore not contributing to the radio flux.

\section{Is VVV\,CL177 a real cluster?}
\repII{Despite of our efforts, there is still no sufficient evidence that VVV\,CL177 is a real cluster. We instead offer two scenarios.

\subsection{Scenario 1: It is not a single very distant cluster}
In that scenario, all of our observations are the result of a coincidental alignment of many components, as suggested by \citet{yamauchi16}. This would explain the lack of coherence in the stars PMs. The group of stars that this study focused on share the same ($H$$-$$K_S$) color because of the extinction of a foreground object. Finally, the stars\,1 and 2 are YSOs associated with another star forming region than the one at the source of the water maser emission.

Would that scenario be the right one, at least in part, we are still confident that our data show that the 8$\mu m$ nebular emission and the $^{13}$CO emission are associated with the water maser.

\subsection{Scenario 2: It is a single very distant cluster}
Our data indicate that the group of stars that share the same ($H$$-$$K_S$) color and that are near the water maser coordinates are not likely clustered. But it remains possible that some of the stars are associated with the star forming region at the source of the maser emission.
}
When we put all the pieces together, we have:
\begin{itemize}
    \item A water maser source 20.4$^{+2.8}_{-2.2}$\,kpc away from the Sun and associated with \object{GAL 007.47+00.06} \citep{Sa17}.
    \item The RDP corresponding to the 8\,$\mu$m emission profile, giving a radius between 0.45$\arcmin$ and $\sim$1$\arcmin$.
    \item The confirmation of the YSO nature of at least one of the redder stars (star\,1), \repII{a few arcsecs from the maser}, plus a reliable YSO candidate (star\,2). The spectra of both objects resemble the \citet{cooper13} massive YSO spectra. 
    \item The RV of star\,1 corresponds to that of the maser, and the $^{13}$CO emission at that velocity corresponds to the 8$\mu$m emission.
    \item An extreme extinction near A$_{\rm V}=38.0\pm$2.5\,mag \repII{determined from} the 1-0 S(1) and Q(3) emission lines of H$_2$ flux ratio for star\,1.
\end{itemize}{}

\repII{In that scenario, we could detect only a few of the brightest stars of VVV\,CL177. The other stars are foreground stars, which is quite likely in that line of sight.} With $K_S\sim13$-14\,mag, these stars \repII{would} have absolute magnitudes $-3.5<M_{K_S} < +1.5$. This is comparable to many members of the \citet{cooper13} sample of embedded massive YSOs, which have $-8.5<M_{K_S} < +3$, comparable extinction in many cases, and luminosities L$=10^3$ to $10^5$ L$_\odot$.

\repII{On can note that the extreme extinction of star\,1 is compatible with a large distance. Unfortunately, the VVV high-resolution 3D extinction maps \citep[][]{Ch13} do not reach a distance of 20\,kpc, but at 10\,kpc, they give A$_{\rm V}$$\sim$14.6$\pm$1.5\,mag. And at 6\,kpc, one of the distances to \object{GAL\,007.47+0.06} by \citet[][]{wink82}, A$_{\rm V}$$\sim$4.0$\pm$0.7\,mag, which is an order of magnitude off, discarding a closer distance for both \object{GAL\,007.47+0.06} and star\,1.}

\section{Conclusions}\label{Conclusions}

\repII{This study offers a glimpse at what could probably be the environment and the stellar population associated to \object{GAL 007.47+0.06}, the HII region on the far side of the Galaxy. It contains a large nebula that spans over 5 to 10\,pc and forms massive stars in its core, as one can deduce from a water maser emission and the presence of at least one massive YSO. It is surrounded by at least $\sim10^3$\,M$_\odot$ of cold molecular gas in the periphery that has not formed stars yet.

From the analysis  of the available data we can't prove definitely  the existence of stellar cluster placed on the far side of the Galaxy. But, assuming a single star forming event, its age could be as young as $\sim$1\,Myr, as the most luminous star (therefore the likely most massive star) is still a YSO.} 

Follow-up spectroscopy and imaging is required for a \repII{deeper investigation of the region. Monitoring of the flux and spectral variation will also allow the identification of binary stars or massive accretion process associated to massive YSOs. The complex PM distribution is indicating that the considered area is potentially significantly contaminated with stars reddened by a foreground object \citep[like the foreground cloud mentioned by][]{yamauchi16} and that appear to share the same colors as the cluster's stars.}


\acknowledgments
ANC would like to gratefully acknowledge the help from Thomas R. Geballe, for his useful tips and enriching conversations. Based on observations obtained at the international Gemini Observatory, a program of NSF’s NOIRLab, which is managed by the Association of Universities for Research in Astronomy (AURA) under a cooperative agreement with the National Science Foundation. on behalf of the Gemini Observatory partnership: the National Science Foundation (United States), National Research Council (Canada), Agencia Nacional de Investigaci\'{o}n y Desarrollo (Chile), Ministerio de Ciencia, Tecnolog\'{i}a e Innovaci\'{o}n (Argentina), Minist\'{e}rio da Ci\^{e}ncia, Tecnologia, Inova\c{c}\~{o}es e Comunica\c{c}\~{o}es (Brazil), and Korea Astronomy and Space Science Institute (Republic of Korea). Support is provided  by ANID, Millennium Science Initiative ICN12-009 (MAS), the ANID FONDECYT Iniciación 11201161, the FONDECYT Iniciaci\'on project 11171025,  the FONDECYT Regular project 1201490, the CONICYT + PAI ``Concurso Nacional Inserci\'on de Capital Humano Avanzado en la Academia 2017'' project PAI 79170089, and the MINEDUC-UA project, code ANT1855. D.M. is supported by the BASAL Center for Astrophysics and Associated Technologies (CATA) through grant AFB 170002, and by project FONDECYT Regular No. 1170121.

%

\vspace{5mm}
\facilities{Spitzer(IRAC), VISTA(VIRCAM), Green Bank Telescope, Gemini(Flamingos-2)}


\software{astrolib \citep{La93},  
          IRAF \citep{To93}, 
          DAOphot \citep{St94}
          }

\end{document}